\documentclass[aps]{revtex}

\usepackage[T1]{fontenc}
\usepackage{a4}
\usepackage{epsfig}
\usepackage{graphics}
\usepackage{graphicx}
\usepackage{latexsym}

\def\i{\'\i}

\def\be{\begin{equation}}
\def\ee{\end{equation}}
\def\bea{\begin{eqnarray}}
\def\eea{\end{eqnarray}}

\def\<{\langle}
\def\>{\rangle}



\begin{document}

\title{ 
Comparison among HB-inspired algorithms for continuous-spin systems
and gauge fields
}

\author{A.\ Cucchieri, R.B. Frigori, T.\ Mendes 
and A.\ Mihara \footnote{Work presented by A.\ Cucchieri at the
{\it IV Brazilian Meeting on Simulational Physics}
-- Ouro Preto - MG /  Brazil, August 2005.}
}

\address{
Instituto de F\'\i sica de S\~ao Carlos -- 
Universidade de S\~ao Paulo \\
C.P.\ 369, 13560-970, S\~ao Carlos, SP, Brazil
}

\date{September 2005}

\maketitle

\begin{abstract}

We propose a new local algorithm for the thermalization of $n$-vector 
spin models, which can also be used in the numerical simulation of
$SU(N)$ lattice gauge theories. The algorithm combines heat-bath (HB) 
and micro-canonical updates in a single step --- as opposed to the
hybrid overrelaxation method, which alternates between the two kinds
of update steps --- while preserving ergodicity. 
We test our proposed algorithm in the case of the one-dimensional
4-vector spin model and compare its performance with the standard HB 
algorithm and with other HB-inspired algorithms.

\end{abstract}

%%%%%%%%%%%%%%%%%%%%%%%%%%%%%%%%%%%%%%%%%%%%%%%%%%%%%%%%%%%%%%%%%%%%%%%%%%%%

\section{Introduction}

The lattice formulation of quantum field theories provides a first
principles approach to investigate non-perturbative aspects of
high-energy physics. In this formulation the theory becomes equivalent
to a statistical mechanical model, which can be studied numerically
using Monte Carlo simulations (see for example
\cite{Mendes:2003gz,Cucchieri:2003zx} and references therein).
As a consequence, the system considered
evolves according to a Markov process in the so-called Monte Carlo time
and the action-weighted configuration-space average of
the observables is substituted by a time average over successive
(independent) field configurations of the system.
The possibility to use Monte Carlo simulations to study the theory
nonperturbatively is especially important in the case of quantum
chromodynamics, the nonabelian $SU(3)$ gauge theory describing the
strong interactions between hadrons. These simulations are computationally
very demanding and must be done using local thermalization algorithms, 
since global methods (such as cluster algorithms) do not work well in 
this case.

We should notice that, based on the above-mentioned equivalence,
when the continuum limit of the lattice quantum field theory is taken,
the corresponding  statistical mechanical model approaches its
critical point. More precisely, the distance from the critical point is
given by the lattice parameter $\beta$ (directly related to the lattice
bare coupling constant), which corresponds to an inverse
temperature. For nonabelian gauge theories (as a consequence of
asymptotic freedom) the continuum limit is given by $\beta\to\infty$, 
i.e.\ the critical point corresponds to temperature zero. The expected
critical behavior is therefore similar to the one of a two-dimensional
continuous-spin model (e.g.\ the classical Heisenberg or $n$-vector spin
model with $n>2$) or of a one-dimensional spin model, which show a critical 
point only at temperature zero, or $\beta\to\infty$.

The process of obtaining independent field configurations is called
{\em thermalization} and is usually carried out by applying at each link of
the lattice a local algorithm, such as Metropolis or heat bath (HB).
When a critical point is approached,
this process is afflicted
by the well-known phenomenon of {\em critical slowing-down} (CSD)
\cite{Sokal:1989ea}, which increases the correlation
among successive field configurations.
This implies that the {\em integrated
auto-correlation time} $\tau_{int}$
increases as a power of the lattice side $N$. In particular,
for the Metropolis or HB algorithms
one has $\tau_{int} \sim N^2$, i.e.\ the {\em dynamic critical exponent} $z$
is equal to 2. Since statistical Monte-Carlo errors are proportional to
$\sqrt{2 \, \tau_{int}}$, numerical simulations become increasingly inefficient
close to a critical point.
In order to reduce the problem of CSD one can
combine the standard Metroplis and HB algorithms with so-called micro-canonical
updates, allowing larger jumps in the configuration space and therefore
improving the generation of independent samples. This is the idea behind
the so-called hybrid overrelaxation method \cite{Wolff:1992ri}.
In general, adding a few micro-canonical sweeps greatly reduces 
CSD and, correspondly, the computational work.

A modification of the heat-bath algorithm, called overheat-bath (OHB),
was introduced some years ago in Ref.\ \cite{Petronzio:1991gp}.
The basic idea was to incorporate a micro-canonical move directly into
the heat-bath step, thus reducing the computational cost while preserving
the large moves in the configuration space. As it turns out, combining the 
two moves (heat-bath and micro-canonical) in a single step leads to a
significant improvement in performance when compared to the hybrid overrelaxation
method described above, which is based on alternating the two kinds
of update moves. The resulting algorithm was
indeed able to speed up the thermalization process, but, as already
stressed in Ref.\ \cite{Petronzio:1991gp},
it is not clear if it preserves ergodicity
(especially when working at small temperatures).
The OHB algorithm is used today in numerical simulations
\cite{Fiore:2003yw}, usually
combined with other algorithms in order to ensure ergodicity.
In this work we propose a modification of the overheat-bath
algorithm, which we call the modified heat-bath algorithm (MHB),
incorporating a micro-canonical move into the heat-bath step
without compromising ergodicity.
In order to test the MHB algorithm --- and compare its performance
with the standard HB algorithm and the OHB algorithm ---
we consider the $4$-vector spin model on a 1-dimensional lattice.
Note that the model is exactly solvable, which makes it especially
suited for testing the algorithm and comparing results to the
exact solution.

Let us also note that,
due to the isometry between the groups
$O(4)$ and $SU(2)$, it is possible to study the $SU(2)$ 
case with the same algorithm used for the $4$-vector case.
More precisely, the {\em local} update --- i.e.\ the update of a single
spin or gauge field variable while holding the rest of the lattice fixed --- 
of the $SU(2)$ lattice gauge theory is identical to the one for the 4-vector 
model.
This does not mean, however, that the {\em global} update of an $SU(2)$
gauge field configuration can be obtained from the corresponding update of
a 4-vector model. Indeed, the latter update can be done very efficiently
using the Swendsen-Wang-Wolff algorithm, while no such class of algorithms
works well for lattice gauge theories. The method proposed here is
therefore intended mostly for use in simulations of lattice gauge theories
and not of $n$-vector models, although hybrid overrelaxation methods have
also been applied in large-scale simulations of $n$-vector models \cite{Chen}. 
Of course, an efficient
thermalization in the $SU(2)$ case is of great physical interest, since the 
quenched $SU(N_c)$ case (for $N_c \geq 3$) is usually studied applying the
$SU(2)$-embedding technique introduced in Ref.\ \cite{Cabibbo:1982zn}.
The $SU(2)$ embedding is also used for simulating other Lie groups,
such as the $Sp(2)$ and $Sp(3)$ groups, in studies of the
deconfinement phase transition \cite{Holland:2003kg}.
Preliminary results for the $2d$ $SU(2)$ case
have been presented in \cite{Frigori:2005ym}.

%%%%%%%%%%%%%%%%%%%%%%%%%%%%%%%%%%%%%%%%%%%%%%%%%%%%%%%%%%%%%%%%%%%%%%%%%%%%

\section{The $4$-vector spin model and the algorithms}

The $\,4$-vector spin model (on a 1-d lattice) is defined by the
Hamiltonian
\be
{\cal H} \,=\,- \beta\, \sum_{x=1}^{N}\,S_{x} \cdot S_{x+1}
\; ,
\label{eq:calH}
\ee
where the spins $\,S_x\,$ are four-dimensional unit vectors, 
$\,\beta\sim$1/Temperature$\,$, $\,N\,$ is the lattice side
and $\,\cdot\,$ indicates a scalar product.

In the case of a local algorithm, one has to consider the contribution
to the Hamiltonian due to a single spin $\,S_x$. This gives
$\, {\cal H}_{ss} = -\beta \, S_{x} \cdot H_{x} \,+\, \mbox{constant}\,$,
where the ``effective magnetic field'' $\,H_x\,$ is given by
$\,H_x = S_{x-1} + S_{x+1} \,$.
(Here we consider periodic boundary conditions.)
The above equation can also be written as
\be
{\cal H}_{ss} \,=\, -\frac{\beta \, N_x}{2}\; 
  \mbox{Tr} \; S_{x} \; \widetilde{H}_{x}^{\dagger}
\,+\, \mbox{constant}\; ,
\label{eq:Hss2}
\ee
where $\,S_x\,$ and $\,\widetilde{H}_x\,$
are now interpreted as $\,SU(2)\,$ matrices in
the fundamental representation and $N_x = \sqrt{\mbox{det}\,H_x}$.
Note that Eq.\ (\ref{eq:Hss2}) is clearly analogous to the expression
of the single-link action obtained by considering the contribution
of a single link variable to the $\,SU(2)\,$ Wilson action
\cite{Frigori:2005ym}.

Using the single-side action (\ref{eq:Hss2}) and the invariance of
the group measure under group multiplication, one obtains the HB
update \cite{Creutz:1980zw}
\be
S_{x}^{old} \;\rightarrow\; S_{x}^{new} \,=\, U\,\widetilde{H}_{x} \; .
\ee
Here, $U = u_0 \mbox{I} + i\,\sum_{j}\,u_j \sigma_j$ is an
$SU(2)$ matrix, 
$\,I\,$ is the $\,2 \times 2\,$ identity matrix, $\,\sigma_j\,$
are the three Pauli matrices and
$u_0$ is randomly generated according to the distribution  
\be
\sqrt{1 - u_0^2}\,\exp{\left( \,\beta\, N_x\, u_0 \, \right)}\, du_0\; .
\label{eq:distr}
\ee
At the same time,
the vector $\vec{u}$ --- with components $\,u_j\,$ normalized to
$\,\sqrt{1 - u_0^2}\,$ --- points along a uniformly chosen random direction
in three-dimensional space \cite{Cucchieri:2003zx}.

In the hybrid over-relaxed algorithm one does $m$ micro-canonical
(or energy-conserving) update sweeps over the lattice, followed 
by one HB sweep. The micro-canonical update is a local deterministic
transformation applied to the $SU(2)$ matrix $S_x$, which does not change
the value of the Hamiltonian. This is achieved by considering the
update
\be
S_{x}^{old} \;\rightarrow\; S_{x}^{new} \,=\, \widetilde{H}_x\,
\mbox{Tr}\left[ S_{x}^{old}\, \widetilde{H}_x^{\dagger} \right] 
\, - \, S_{x}^{old} \; .
\label{eq:micro}
\ee
As stressed in the Introduction,
this update represents a large move in the configuration space, allowing
the hybrid over-relaxed algorithm to reduce CSD at the
price of a greater computational cost, due to the micro-canonical sweeps.
Usually, by setting $m=2,3$ one obtains a strong reduction in the value
of $\tau_{int}$ while increasing the computational cost by a factor smaller
than 2.

In the OHB \cite{Petronzio:1991gp} one tries to include the micro-canonical
step (\ref{eq:micro}) directly into the heat-bath algorithm. To this end one 
generates $u_0$ according to the distribution (\ref{eq:distr})
while the components of $\vec{u}$ are not randomly chosen but are set
using the relation $\, \vec{u} = -\vec{w}\,$,
where $W = w_0\mbox{I} + i\vec{\sigma}\cdot\vec{w} =
S_x^{old}\,\widetilde{H}_x^{\dagger}$.
As a final step, the vector $\vec{u}$ is
normalized to $\sqrt{1 - u_0^2}$. Clearly, the idea here is the
same one applied in the standard hybrid over-relaxed algorithm: one tries
to maximize the move in the configuration space by changing the
sign of the component of $\,W\,$ that is orthogonal to
the effective magnetic field. (Note that
the action $S = -\beta\,N / 2 \,\mbox{Tr} \, W$
can be viewed as the action of a matrix $W$ in an effective magnetic field
given by the identity matrix $\mbox{I}$.) Clearly, this step does not obey
the uniform distribution for $\vec{u}$ but is a microcanonical move.
Indeed we can think of the algorithm as a two-step process: a HB move
followed by a microcanonical step. Thus, the algorithm is {\em exact}
but may not be ergodic. We analyze the conditions for
applicability of the OHB algorithm in a separate work \cite{future}, but
it is clear that for some initial configurations the algorithm can get
``trapped'' inside a subset of the space of configurations, compromising
the ergodicity condition. One such configuration is, for example, a ``cold''
start for an $n$-vector model, in which all spins are 
aligned along a fixed direction.
In this case, the OHB is not able to change the lattice configuration at all.
There is a clear problem also if the initial configuration in the 
$SU(2)$-lattice-gauge-theory case is given by variables in an Abelian subgroup.
As can be easily seen, the update will change the initial configuration in
this case, but the resulting Markov chain will remain restricted to the Abelian
sector, without exploring the full space of configurations.

In this work we propose a modified HB algorithm (MHB) in which the generation of
the $SU(2)$ matrix $U$ is done as in the HB case, but followed by
one final step: \textit{if the scalar product
$\vec{u} \cdot \vec{w}$ is positive then
one does $\vec{u} \rightarrow -\vec{u}$}, where the matrix
$ \, W\,$ has been defined above.
This may also be thought of as a modification of the overheat-bath (OHB).
As said above, the basic idea, in both cases, is to incorporate a
micro-canonical move into the heat-bath step.
The difference is that in our case
the direction of $\vec{u}$ is randomly chosen and only its
sign is modified, according to the above rule, while in the OHB
algorithm one fixes $\vec{u} \propto - \vec{w}$ directly. Our modification
should ensure ergodicity while keeping the advantage of having a
micro-canonical step included into the HB step. We have indeed verified
that the MHB algorithm has no problem with cold starts and shows
a better $\,\tau_{int}\,$ than that of the OHB algorithm
for energy-related observables.
On the other hand, in the OHB case, the move in configuration space
is more optimized and the iteration cost is lower, since the algorithm
is simpler and it needs fewer random numbers.
Our results are presented in the next section.

%%%%%%%%%%%%%%%%%%%%%%%%%%%%%%%%%%%%%%%%%%%%%%%%%%%%%%%%%%%%%%%%%%%%%%%%%%%%

\section{Results and Conclusions}

In order to study the CSD of the new algorithm we have to investigate if,
and with what exponent, its relaxation time $\,\tau_{int}\,$
(for certain quantities we are interested in)
diverges as the lattice size $\,N\,$ increases. To this end,
we have to evaluate $\,\tau_{int}\,$ for different lattice sides $\,N\,$
at ``constant
physics'', namely as the lattice size is increased, the physical size of the
lattice should remain fixed. This is done by introducing a correlation length
$\,\xi\,$ and by keeping the ratio $\,\xi / N\,$ fixed, with $\,N \ll \xi\,$
in order to keep finite-size effects under control.
For the 1-d $4$-vector spin model one has $\,\xi \propto \beta\,$
\cite{Cucchieri:1995yd}, thus one should tune $\,N \propto \beta\,$ in order
to keep the ratio $\,\xi / N\,$ constant.
In our simulations we considered the lattice sizes
$\,N = 32$, $64$, $96$, $128$, $160$, $224$, $256\,$ and $\,\beta =
 2.5 \, N / 32\,$ and $\,\beta = 5.0\, N / 32 $.
In all cases we did $\,1.1 \times 10^6\,$ thermalization sweeps and for the
analysis we discarded the first $\,10^5\,$ sweeps.

In order to test the new algorithm we have evaluated the energy density
$\, \epsilon = N^{-1}\,\sum_{x=1}^{N}\,S_x \cdot S_{x+1}\,$,
the magnetization $\, {\cal M}^2 = (\, 
\sum_{x=1}^{N}\,S_{x}\,)^2\,$
and the magnetic susceptibility $\, \chi = N^{-1}\, \< {\cal M}^2 \> \,$.
Let us recall that there exists an exact solution \cite{Cucchieri:1995yd} for
the 1-d $4$-vector spin model,
even at finite lattice side $\,N\,$ and with periodic
boundary conditions. Thus, we can check the numerical
results for the various quantities
against the exact solution
(see also \cite[Table 1]{Mendes:1995jh}).

In all cases we have evaluated the integrated auto-correlation time
$\tau_{int}$ using an automatic windowing procedure \cite{Sokal:1989ea}
with two different window factors ($c = 4$ and $8$). We also employ a method
based on a comparison between
the naive statistical error with a binning error \cite{deDivitiis:1994yz}.
We checked that these three estimates are always in agreement.
The error on the integrated auto-correlation time
$\tau_{int}$ has been evaluated using the Madras-Sokal formula
\cite{Sokal:1989ea}.

\begin{figure}
\begin{center}
\includegraphics[height=0.40\hsize]{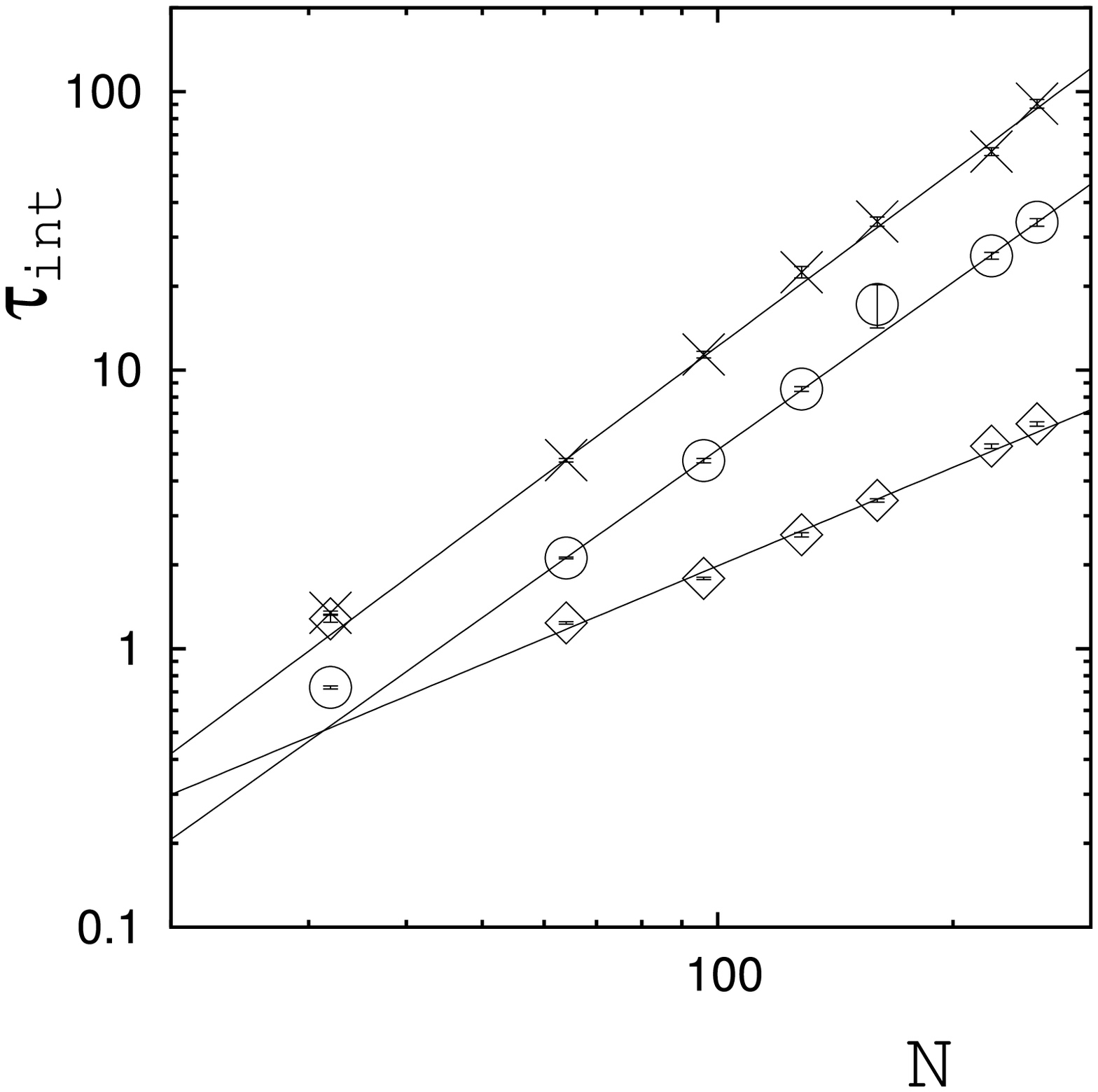}
\protect\hskip 1cm
\includegraphics[height=0.40\hsize]{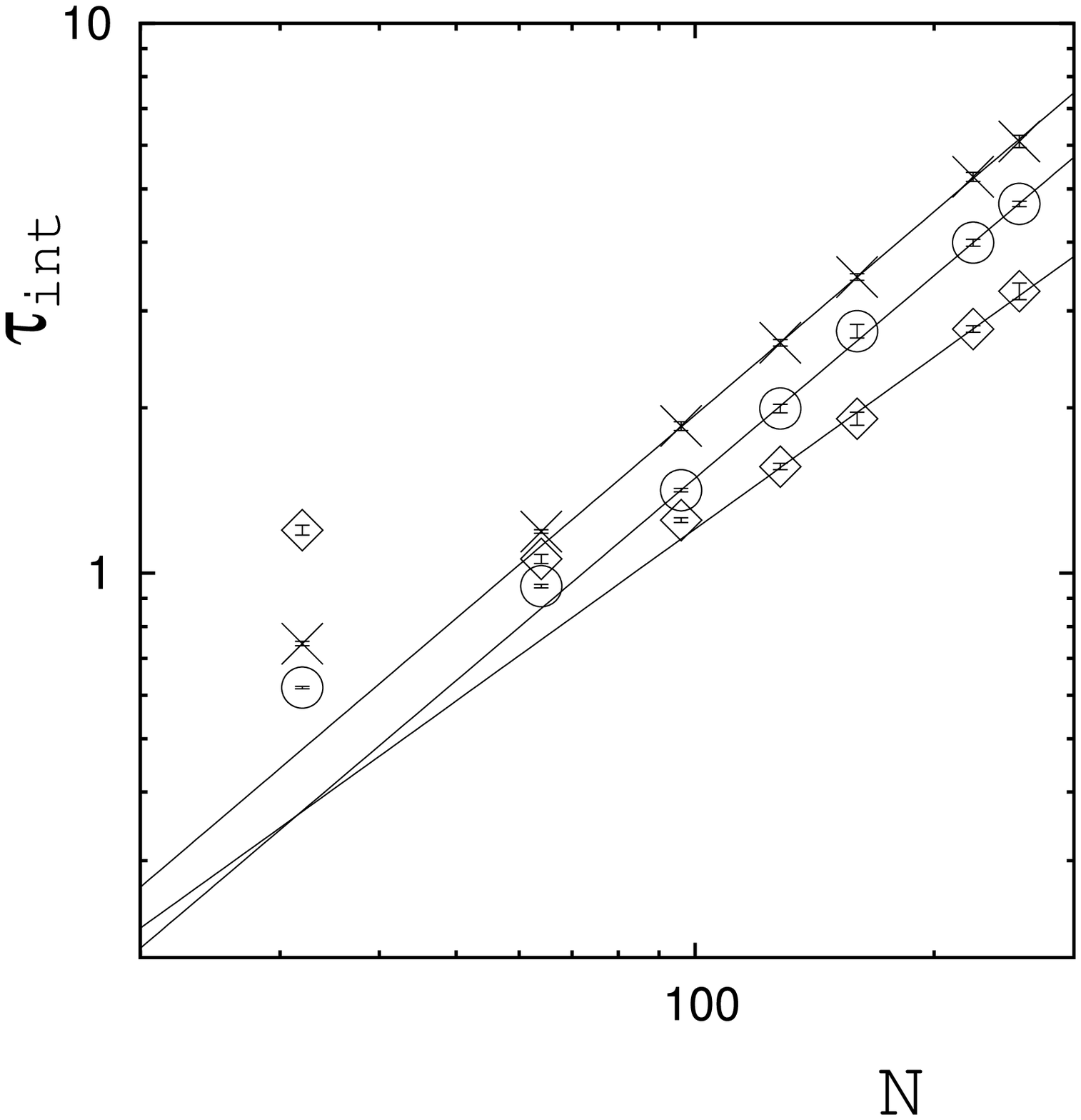}  
\protect\vskip 0.4cm
\caption{Integrated auto-correlation time of the susceptibility
as a function of the lattice side $N$
for HB ($\times$), OHB ($\diamond$) and MHB ($\circ$), with 0 (left)
and 1 (right) overrelaxation step. Data have been fitted to the
Ansatz $\,\tau_{int} = a N^z$.
\label{fig1} }
\end{center}
\end{figure}

\begin{figure}
\begin{center}
\includegraphics[height=0.40\hsize]{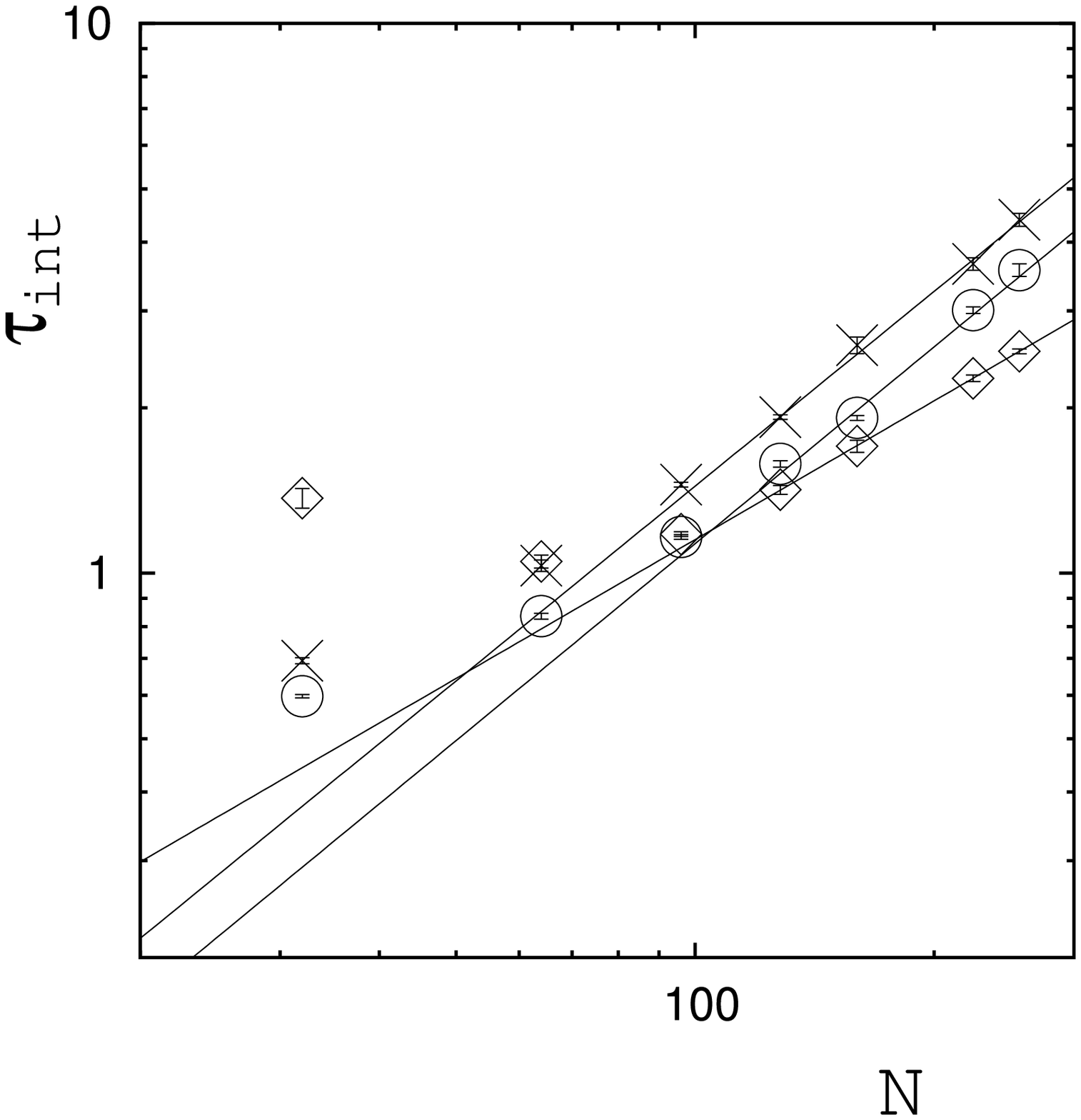}  
\protect\hskip 1cm
\includegraphics[height=0.40\hsize]{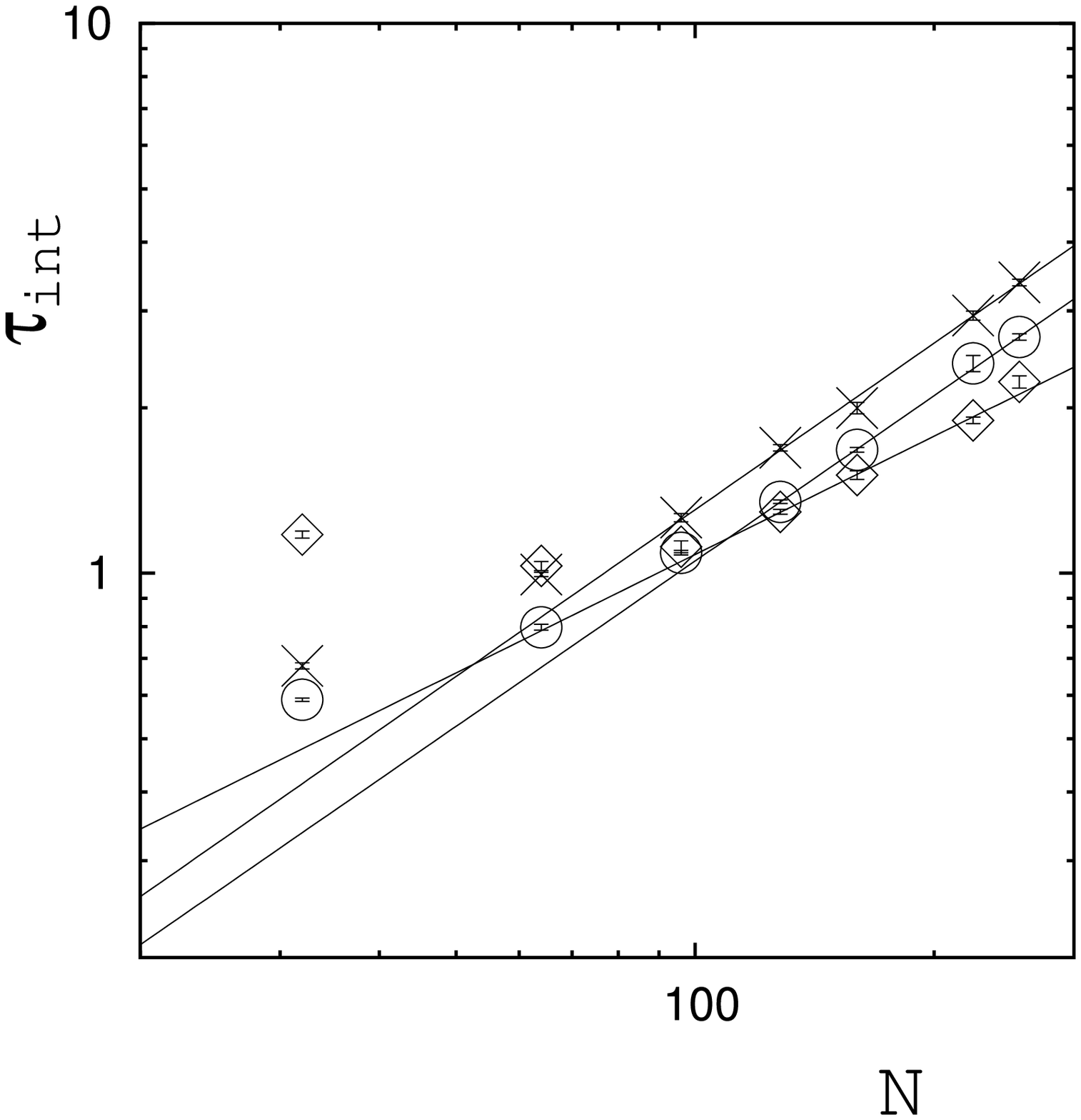}  
\protect\vskip 0.4cm
\caption{Integrated auto-correlation time of the susceptibility
as a function of the lattice side $N$
for HB ($\times$), OHB ($\diamond$) and MHB ($\circ$), with 2 (left)
and 3 (right) overrelaxation step. Data have been fitted to the
Ansatz $\,\tau_{int} = a N^z$.
\label{fig2}}
\end{center}
\end{figure}

Results are reported in Figs.\ \ref{fig1} and \ref{fig2}.
The data obtained for
$\tau_{int}$ have been fitted to the Ansatz $\,\tau_{int} = a \,N^z\,$
in order to estimate the dynamic critical exponent $\,z$.
Our study shows that the MHB algorithm has essentially the same
critical exponent $\,z\,$ of the standard HB algorithm but the
value of the integrated auto-correlation time
$\,\tau_{int}\,$ is about $30 \%$ smaller
compared to the standard HB algorithm.
This implies a reduction in the statistical error and in the
computational cost by a factor of about $20\%$.
Similar results have been obtained in the $SU(2)$ case
\cite{Frigori:2005ym}.
Thus, a good decorrelation can be reached without the use of multiple
micro-canonical sweeps. In particular, from our simulations
it appears that the HB algorithm with $m$ micro-canonical steps
is essentially equivalent to the MHB algorithm with $m-1$
micro-canonical steps. This feature may be
useful in parallel computing \cite{Cucchieri:2003zx}
due to the reduction of the inter-node communication.

At the same time, from our data one sees that
the MHB algorithm has a critical exponent
$z$ larger than that of the OHB algorithm.
In order to reduce CSD for the MHB algorithm,
while keeping its ergodicity properties, we are now testing
an algorithm that interpolates between MHB and OHB.

A more extensive study of these algorithms, both in the $O(4)$ and
in the $SU(2)$ cases, will be presented in a forthcoming work
\cite{future}.

%%%%%%%%%%%%%%%%%%%%%%%%%%%%%%%%%%%%%%%%%%%%%%%%%%%%%%%%%%%%%%%%%%%%%%%%%%%%

\section*{Acknowledgments}

Research supported by FAPESP (Projects No.\
00/05047-5 and 03/00928-1). Partial support
from CNPq is also acknowledged  (AC, TM).

%%%%%%%%%%%%%%%%%%%%%%%%%%%%%%%%%%%%%%%%%%%%%%%%%%%%%%%%%%%%%%%%%%%%%%%%%%%%

%%%%%%%%%%%%%%%%%%%%%%%%%%%%%%%%%%%%%%%%%%%%%%%%%%%%%%%%%%%%%%%%%%%%%%%%%%%%

\end{document}